\documentclass[letterpaper, 10 pt, conference]{ieeeconf}  
\IEEEoverridecommandlockouts
\usepackage{balance}
\usepackage{color}
\usepackage{caption}
\usepackage{subcaption}
\usepackage{enumerate}
\usepackage{cite}
\usepackage{empheq}
\usepackage{mathrsfs}

\usepackage{mathtools}

\usepackage{tikz}
\usepackage{circuitikz}


\makeatletter
\pgfcircdeclarebipole{}{\ctikzvalof{bipoles/interr/height 2}}{spst}{\ctikzvalof{bipoles/interr/height}}{\ctikzvalof{bipoles/interr/width}}{

    \pgfsetlinewidth{\pgfkeysvalueof{/tikz/circuitikz/bipoles/thickness}\pgfstartlinewidth}

    \pgfpathmoveto{\pgfpoint{\pgf@circ@res@left}{0pt}}
    \pgfpathlineto{\pgfpoint{.6\pgf@circ@res@right}{\pgf@circ@res@up}}
    \pgfusepath{draw}   
}

\def\pgf@circ@spst@path#1{\pgf@circ@bipole@path{spst}{#1}}
\tikzset{switch/.style = {\circuitikzbasekey, /tikz/to path=\pgf@circ@spst@path, l=#1}}
\tikzset{spst/.style = {switch = #1}}
\makeatother

\makeatletter
\let\proof\@undefined                        
\let\endproof\@undefined                  
\makeatother
\usepackage{graphicx,amssymb,amstext,amsmath,amsthm}

\usepackage[bookmarks=true]{hyperref}
\usepackage{algorithm,algorithmicx,algpseudocode}
\algnewcommand{\algorithmicgoto}{\textbf{go to}}%
\algnewcommand{\Goto}[1]{\algorithmicgoto~\ref{#1}}%
\algnewcommand{\LineComment}[1]{\Statex \(\triangleright\) #1}
\algnewcommand{\LineCommentN}[1]{\Statex \hspace{1cm}\(\triangleright\) #1}

\usepackage{multirow}
\usepackage{stfloats}

\newtheorem{prop}{Proposition} 
\newtheorem{cor}{Corollary}
\newtheorem{thm}{Theorem}
	\newtheorem{assumption}{Assumption}
\newtheorem{lem}{Lemma}
\newtheorem{defn}{Definition}

\newtheorem{problem}{Problem}

\usepackage{setspace}

\let\oldbibliography\thebibliography
\renewcommand{\thebibliography}[1]{%
  \oldbibliography{#1}%
}

\newcommand{\yong}[1]{{\color{black} #1}}
\newcommand{\moh}[1]{{\color{black} #1}}
\newcommand{\yongn}[1]{{\color{black} #1}}

\begin{document}

\title{\LARGE \bf Simultaneous Input and State Interval Observers\\ for Nonlinear Systems \moh{with Full-Rank Direct Feedthrough}} 

\author{%
Mohammad Khajenejad, Sze Zheng Yong\\
\thanks{
M. Khajenejad and S.Z. Yong are with the School for Engineering of Matter, Transport and Energy, Arizona State University, Tempe, AZ, USA (e-mail: \{mkhajene, szyong\}@asu.edu).}
\thanks{This work is partially supported by NSF grant CNS-1932066.}
}

\maketitle
\thispagestyle{empty}
\pagestyle{empty}

\begin{abstract}
A simultaneous input and state interval observer is presented for 
Lipschitz \yong{continuous} nonlinear systems with unknown inputs and bounded noise signals \yong{for the case when the direct feedthrough matrix has full column rank}. The observer leverages the existence of bounding decomposition functions for mixed monotone mappings to recursively compute the maximal and minimal elements of the estimate intervals that are compatible with output/measurement signals, and are proven to contain the true state and unknown input. 
Furthermore, we derive a Lipschitz-like property for decomposition functions, which provides several sufficient conditions for stability of the designed observer and boundedness of the sequence of estimate interval widths. 
Finally, the effectiveness of our approach is demonstrated using an illustrative example.
 \end{abstract}
\section{Introduction}
\emph{Motivation.} \moh{State and unknown input estimation has \yong{recently emerged as an important and indispensable component in many engineering applications} 
such as fault detection, urban transportation, aircraft tracking and attack (unknown input) detection and mitigation in cyber-physical systems \cite{liu2011robust,yong2016tcps,yong2018simultaneous}. Particularly, in bounded-error settings, interval/set-membership approaches have been proposed to provide hard accuracy bounds, which \yong{is especially useful for safety-critical systems \cite{blanchini2012convex}.} 
Moreover, \yong{since the unknown inputs may be strategic in adversarial settings,} 
\yong{the ability to} simultaneously estimate states and inputs without imposing any assumption on the unknown inputs} \yong{is desirable and often crucial}.

\emph{Literature review.} 
Several approaches have been proposed in the literature to design interval observers  \cite{jaulin2002nonlinear,kieffer2004guaranteed,moisan2007near,bernard2004closed,raissi2010interval,raissi2011interval,mazenc2011interval,mazenc2013robust,wang2015interval,efimov2013interval,zheng2016design}. However, these approaches often hinge upon relatively strong assumptions about the existence of certain system properties, such as monotone dynamics, \cite{moisan2007near,bernard2004closed}, Metzler and/or Hurwitz partial linearization of nonlinearities \cite{raissi2011interval,mazenc2013robust}, cooperativeness \cite{raissi2010interval}, linear time-invariant (LTI) dynamics \cite{mazenc2011interval} and linear parameter-varying (LPV) dynamics that admits a diagonal Lyapunov function \cite{wang2015interval}. 
Moreover, the work in \cite{efimov2013interval} addresses the design of interval observers for a class of continuous time nonlinear systems without unknown inputs using bounding functions by imposing somewhat restrictive assumptions on the nonlinear dynamics to conclude stability, without discussing necessary and/or sufficient conditions for the existence of bounding functions or how to compute them. The authors in \cite{zheng2016design} study the problem of interval state estimation for a class of uncertain nonlinear systems, by extracting a known nominal observable subsystem from the plant equations and designing the observer for the transformed system. However, the derived conditions for the existence and stability of the observer is not \emph{constructive}. 
Moreover, there is no guarantee that the derived functional bounds have finite values, i.e., be bounded sequences. More importantly, the aforementioned works do not consider unknown inputs (different from bounded-norm noise/disturbance) nor the reconstruction/estimation of the uncertain inputs.

The problem of designing an unknown input interval observer that satisfies $L_2/L_{\infty}$ optimality criteria is investigated in \cite{ellero2019unknown} where the required Metzler property is formulated as a part of a semi-definite program. However, unfortunately, their approach is limited to continuous-time LPV systems. Moreover, in their setting, the (potentially unbounded) unknown inputs do not affect the output (measurement) equation. On the other hand, for systems with linear output equations and where both the state and output equations are affected by unknown inputs/attacks, the problem of simultaneously designing state and unknown input set-valued observers  has been studied in our previous works for LTI \cite{yong2018simultaneous}, LPV \cite{khajenejad2019simultaneous} and switched linear \cite{khajenejadasimultaneous} systems with bounded-norm noise.       

 \emph{Contributions.} 
 \moh{We consider} \yong{the design 
 of} {an} observer that \moh{\emph{simultaneously} returns} interval-valued estimates of states and unknown inputs for \moh{a broad range of nonlinear} systems. \moh{Our approach is novel in multiple \yong{ways}. First, to the best of our knowledge, all \yong{existing} 
  interval observers in the literature only return either state \cite{jaulin2002nonlinear,kieffer2004guaranteed,moisan2007near,bernard2004closed,raissi2010interval,raissi2011interval,mazenc2011interval,mazenc2013robust,wang2015interval,efimov2013interval,zheng2016design} or input \cite{ellero2019unknown} estimates, whereas our \yong{observer} simultaneously returns both. Second, we consider}  
  arbitrary \yong{unknown} input signals, 
  \moh{i.e., no restrictive assumptions, such as being bounded or stochastic with zero mean \yong{(as is often assumed for noise)}, are imposed on \yong{the} unknown inputs. Third,} leveraging decomposition functions as \emph{\moh{nonlinear} bounding mappings} of mixed monotone vector fields \cite{yang2019sufficient,coogan2015efficient}, \moh{which include almost every realistic nonlinear function \yong{\cite{yang2019tight}}, we show that our interval estimates are compatible with measurement outputs} 
  and are guaranteed to contain the true states and unknown inputs. 
 \moh{Fourth}, we provide \moh{several} sufficient conditions \moh{in the form of Linear Matrix Inequalities (LMI)} for the stability of our designed observer.
  \moh{\yong{Finally,} we provide upper bounds for the interval widths at each time step, as well as their \yong{steady-state} values}. 
\section{Preliminaries}

{\emph{Notation.}} $\mathbb{R}^n$ denotes the $n$-dimensional Euclidean space and $\mathbb{R}_{+{+}}$ positive real numbers. 
For vectors $v,w \in \mathbb{R}^n$ and a matrix $M \in \mathbb{R}^{p \times q}$, $\|v\|\triangleq \sqrt{v^\top v}$ and $\|M\|$ denote their (induced) 2-norm, \yongn{and} 
$v \leq w$ is an element-wise inequality. 
Moreover, the transpose, 
Moore-Penrose pseudoinverse{, $(i,j)$-th element} 
and rank of $M$ are given by $M^\top$, 
$M^\dagger${, $M_{i,j}$} 
and ${\rm rk}(M)$. { We call $M$ a non-negative matrix, i.e., $M \geq 0$, if $M_{i,j} \geq 0, \forall i \in \{1\dots p\},\forall j \in \{1 \dots q \}$}. For a symmetric matrix $S$, $S \succ 0$ and $S \prec 0$ ($S \succeq 0$ and $S \preceq 0$) are
 positive and negative (semi-)definite, respectively.

Next, we introduce some definitions and related results that will be useful throughout the paper.

\begin{defn}[Interval, Maximal and Minimal Elements, Interval Width]\label{defn:interval}
Set $\mathcal{I} \in \mathbb{R}^n$ is called an interval in $\mathbb{R}^n$, if $\exists \underline{s},\overline{s} \in \mathcal{I}$ such that $\underline{s} \leq x \leq \overline{s}, \forall x \in \mathcal{I}$. $\underline{s}$, $\overline{s}$ and $\|\overline{s}-\underline{s}\|$ are called the minimal element, the maximal element and the {width} of $\mathcal{I}$, respectively. 
\end{defn}
\begin{prop}\cite[Lemma 1]{efimov2013interval}\label{prop:bounding}
Suppose $\underline{b} \leq b \leq \overline{b}$, where $\underline{b}, b, \overline{b} \in \mathbb{R}^n$. Let $A \in \mathbb{R}^{m \times n}$. Then\moh{,} $A^+\underline{b}-A^{++}\overline{b} \leq Ab \leq A^+\overline{b}-A^{++}\underline{b}$, where $A^+,A^{++} \in \mathbb{R}^{m \times n}$, $A^+_{i,j}=A_{i,j}$ if $A_{i,j} \geq 0$, $A^+_{i,j}=0$ if $A_{i,j} < 0$ and $A^{++}=A^+-A$.
\end{prop}
\begin{cor} \label{cor:bounding}
If $A \in \mathbb{R}^{n \times m}$ is a non-negative matrix (element-wise), then $A\underline{b} \leq Ab \leq A\overline{b}$. 
\end{cor}
\begin{defn}[Lipschitz Continuity]\label{defn:lip}
Vector field $f(\cdot):\mathcal{D}_f \rightarrow \mathbb{R}^m$ is globally $L_f$-Lipschitz continuous on $\mathcal{D}_f \subseteq \mathbb{R}^n$, if there exists $L_f \in \mathbb{R}_{+{+}}$, such that $\|f(x_1)-f(x_2)\| \leq L_f \|x_1-x_2\|$, $ \forall x_1,x_2 \in D_f$.
\end{defn}
\begin{defn}[Mixed-Monotone Mappings and Decomposition Functions]\cite[Definition 4]{yang2019sufficient}\label{defn:mixed-monotone}
A mapping $f:\mathcal{X} \subseteq \mathbb{R}^n \rightarrow \mathcal{T} \subseteq \mathbb{R}^m$ is mixed monotone if there exists $f_d:\mathcal{X} \times \mathcal{X} \rightarrow \mathcal{T}$ satisfying the following:
\begin{enumerate}
\item $f$ is embedded on the diagonal of $f_d$, i.e., $f_d(x,x)=f(x)$,  
\item $f_d$ is monotone increasing in its first argument, i.e., $x_1 \geq x_2 \implies f_d(x_1,y)\geq f_d(x_2,y)$, and
\item $f_d$ is monotone decreasing in its second argument, i.e., $y_1 \geq y_2 \implies f_d(x,y_1) \leq f_d(x,y_2)$.
\end{enumerate} 
A function $f_d$ satisfying the above conditions is called a decomposition function of $f$. 
\end{defn}
\begin{prop}\cite[Theorem 1]{coogan2015efficient}\label{prop:embedding}
Let $f:\mathcal{X} \subseteq \mathbb{R}^n \rightarrow \mathcal{T} \subseteq \mathbb{R}^m$ be a mixed monotone mapping with decomposition function $f_d:\mathcal{X} \times \mathcal{X} \rightarrow \mathcal{T}$ and $\underline{x} \leq x \leq \overline{x}$, where $\underline{x},x,\overline{x} \in \mathcal{X}$. Then $f_d(\underline{x},\overline{x}) \leq f(x) \leq f_d(\overline{x},\underline{x})$.
\end{prop}

\section{Problem Formulation} \label{sec:Problem}
\noindent\textbf{\emph{System Assumptions.}} 
Consider the nonlinear discrete-time system with unknown inputs and bounded noise 
\begin{align} \label{eq:system}
\begin{array}{ll}
x_{k+1}&=f(x_k)+B u_k+G d_k + w_k,\\
y_k&=g(x_k) +D u_k + H d_k + v_k, \end{array}
\end{align}
where $x_k \in \mathbb{R}^n$ is the state vector at time $k \in \mathbb{N}$, 
$u_k \in \mathbb{R}^m$ is a known input vector, $d_k \in \mathbb{R}^p$ is an unknown input vector, and $y_k \in \mathbb{R}^l$ is the measurement vector. The process noise $w_k \in \mathbb{R}^n$ and the measurement noise $v_k \in \mathbb{R}^l$ are assumed to be bounded, with $\underline{w} \leq w_k \leq \overline{w}$ and $\underline{v} \leq v_k \leq \overline{v}$, where $\underline{w}$, $\overline{w}$ and $\underline{v}$, $\overline{v}$ are the known \yongn{lower  and upper bounds of the process and measurement noise signals, respectively.} 
We also assume \yongn{that} lower and upper bounds, $\underline{x}_0$ and $\overline{x}_0$, for the initial state ${x}_0$ {are} available, i.e., $\underline{x}_0 \leq x_0 \leq \overline{x}_0$. The vector fields $f(\cdot):\mathbb{R}^n \rightarrow \mathbb{R}^n$, $g(\cdot):\mathbb{R}^n \rightarrow \mathbb{R}^l$ and matrices $B$, $D$, $G$ and $H$ are known 
and of appropriate dimensions, where $G$ and $H$ are matrices that 
encode the \emph{locations} through which the unknown input \yongn{(or attack)} signal can affect the system dynamics and measurements. 
Without loss of generality, we assume that ${\rm rk}[G^\top\; H^\top ]=p$, $n \geq l \geq 1$, $l \geq p \geq 0$ and $m \geq 0$. Moreover, we assume 
the following: 
\begin{assumption}\label{assumption:Hfull}
The direct feedthrough 
matrix $H$ has full column rank.
\end{assumption}
\begin{assumption}\label{assumption:mix-monotone}
Vector fields $f(\cdot)$ and $g(\cdot)$ are mixed-monotone with decomposition functions $f_d(\cdot,\cdot):\mathbb{R}^{n \times n} \rightarrow \mathbb{R}^n$ and $g_d(\cdot,\cdot):\mathbb{R}^{n \times n} \rightarrow \mathbb{R}^l$, respectively.
\end{assumption}
\begin{assumption}\label{assumption:mix-lip}
Vector fields $f(\cdot)$ and $g(\cdot)$ are globally $L_f$-Lipschitz and $L_g$-Lipschitz continuous, respectively. 
\end{assumption}
\yongn{Note} that \moh{Assumption \ref{assumption:Hfull} is a 
common assumption in the \yong{unknown input} observer design literature, \yong{e.g., \cite{gillijns2007unbiased}}}, 
\yongn{while} Assumption \ref{assumption:mix-monotone} is satisfied for a broad range of nonlinear \yongn{functions \cite{yang2019tight}}. 
\yongn{Moreover,} the decomposition function of a 
vector field 
\yongn{is not unique and a specific one is given} 
in \cite[Theorem 2]{yang2019sufficient}\yongn{: If} a vector field $h=\begin{bmatrix} h^\top_1 & \dots & h^\top_n \end{bmatrix}^\top:X \subseteq \mathbb{R}^n \rightarrow \mathbb{R}^m$ is differentiable and its partial derivatives are bounded with known bounds, i.e., $\frac{\partial h_i}{\partial x_j} \in (a^h_{i,j},b^h_{i,j}), \forall x \in X \in \mathbb{R}^n$, where $a^h_{i,j},b^h_{i,j} \in \overline{\mathbb{R}}$, then $h$ is mixed monotone with a decomposition function $h_d=\begin{bmatrix} h^\top_{d1} & \dots & h^\top_{di} & \dots h^\top_{dn} \end{bmatrix}^\top$, where $h_{di}(x,y)=h_{i}(z)+(\alpha^h_i-\beta^h_i)^\top (x-y), \forall i \in \{1,\dots,n\}$, and $z,\alpha^h_i,\beta^h_i \in \mathbb{R}^n$ can be computed in terms of $x, y, a^h_{i,j}, b^h_{i,j}$ as given in \cite[(10)--(13)]{yang2019sufficient}. Consequently, for $x=[x_1\dots x_j \dots x_n]^\top$, $y=[y_1\dots y_j \dots y_n]^\top$, we have 
\begin{align}\label{eq:decompconstruct}
h_{d}(x,y)&=h(z)+C_h(x-y),
\end{align}
with $C_h \triangleq \begin{bmatrix} [\alpha^h_1-\beta^h_1] & \hspace{-0.1cm}\dots \hspace{-0.1cm} & [\alpha^h_i-\beta^h_i] & \dots [\alpha^f_m-\beta^f_m] \end{bmatrix}^\top \in \mathbb{R}^{m \times n} $,  
$\alpha^f_i,\beta^f_i$ as given in \cite[(10)--(13)]{yang2019sufficient}, $z=[z_1 \dots z_j \dots z_m]^\top$ and $z_j=x_j$ or $y_j$ (dependent on the case, cf. \cite[Theorem 1 and (10)--(13)]{yang2019sufficient} for details). On the other hand, when the precise lower and upper bounds, $a_{i,j}, b_{i,j}$, of the partial derivatives are not known or are hard to compute, we can obtain upper and lower approximations of the bounds by using 
\emph{affine abstraction} algorithms, e.g., \cite[Theorem 1]{singh2018mesh}, with 
the slopes \yong{set} to zero.\\[-0.3cm] 

\noindent \textbf{\emph{Unknown Input (or Attack) Signal Assumptions.}} 
The unknown inputs $d_k$ are not constrained to be a signal of any type (random or strategic) nor to follow any model, thus no prior `useful' knowledge of the dynamics of $d_k$ is available (independent of $\{d_\ell\}$ $\forall k\neq \ell$, $\{w_\ell\}$ and $\{v_\ell\}$ $ \forall  \ell$). We also do not assume that $d_k$ is bounded or has known bounds and thus, $d_k$ is suitable for representing adversarial 
attack signals.\\[-0.3cm]

The 
observer design problem 
 can be stated as follows:
\begin{problem}\label{prob:SISIO}
Given a nonlinear discrete-time system with unknown inputs and bounded noise \eqref{eq:system}, design a stable observer that simultaneously finds bounded intervals 
of compatible states and unknown inputs. 
\end{problem}

\section{Simultaneous Input and State Interval Observers (SISIO)} \label{sec:observer}
\subsection{Interval Observer Design} \label{sec:obsv}
We consider a recursive two-step interval-valued observer design, composed of a \emph{state estimation} step and an \emph{unknown input estimation} step with the following form:  
\begin{align*}
\text{\emph{State Estimation:}}  \ \ \mathcal{I}^x_{k} &= \mathcal{F}_x(\mathcal{I}^x_{k-1},\mathcal{I}^d_{k-1},u_{k-1}),\\
\text{\emph{Unknown Input Estimation:}}  \ \mathcal{I}^d_{k}  &= \mathcal{F}_d(\mathcal{I}^x_{k},y_k,u_k),
\end{align*}
where $\mathcal{F}_x$ and $\mathcal{F}_d$ are the to-be-designed interval mappings, while $\mathcal{I}^x_{k}$ and $\mathcal{I}^d_{k}$ are the intervals of compatible states and unknown inputs at time $k$ of the form: 
\begin{align*}
\mathcal{I}^d_{k}&=\{d \in \mathbb{R}^p: \underline{d}_k \leq d \leq \overline{d}_k \},\\
\mathcal{I}^x_{k}&=\{x \in \mathbb{R}^n: \underline{x}_{k} \leq x \leq \overline{x}_{k}\},
\end{align*}
i.e., we restrict the estimation errors to closed intervals in the Euclidean space. In this case, the observer design problem boils down to finding the minimal and maximal elements 
$\underline{d}_k$, $\overline{d}_k$, $\underline{x}_{k}$ and $\overline{x}_{k}$ of the intervals $\mathcal{I}^d_{k}$ and $\mathcal{I}^x_{k}$. 
Our interval observer can be defined at each time step $k \geq 1$ as follows (with known $\underline{x}_{0}$ and $\overline{x}_0$ such that $\underline{x}_{0} \leq x_0 \leq \overline{x}_0$):\\[-0.2cm]

\noindent \emph{State Estimation}: \vspace{-0.1cm}
\begin{align}
  \overline{x}_{k}&\hspace{-0.1cm}= \hspace{-0.1cm}f_d(\overline{x}_{k-1},\underline{x}_{k-1})\hspace{-0.1cm}+\hspace{-0.1cm}Bu_{k-1}\hspace{-0.1cm}+\hspace{-0.1cm}G^+\overline{d}_{k-1}\hspace{-0.1cm}-\hspace{-0.1cm}G^{++}\underline{d}_{k-1}\hspace{-0.1cm}+\hspace{-0.1cm}\overline{w},\label{eq:xup}
\\ \underline{x}_{k}&\hspace{-0.1cm}=\hspace{-0.1cm} f_d(\underline{x}_{k-1},\overline{x}_{k-1})\hspace{-0.1cm}+\hspace{-0.1cm}Bu_{k-1}\hspace{-0.1cm}+\hspace{-0.1cm}G^+\underline{d}_{k-1}\hspace{-0.1cm}-\hspace{-0.1cm}G^{++}\overline{d}_{k-1}\hspace{-0.1cm}+\hspace{-0.1cm}\underline{w}.\label{eq:xlow}
\end{align}
\noindent \emph{Unknown Input Estimation}: \vspace{-0.1cm}
\begin{align}
\overline{d}_{k}&=\min (\overline{d}^1_{k} , \overline{d}^2_{k}), \quad \underline{d}_{k}=\max (\underline{d}^1_{k},\underline{d}^2_{k}), \label{eq:dup}
\end{align}
where
\begin{align}
\overline{d}^1_{k} &=J^+ \overline{r}_k-J^{++}\underline{r}_k  \quad , \quad \underline{d}^1_{k} =J^+ \underline{r}_k-J^{++}\overline{r}_k, \label{eq:dup1} \\
\overline{d}^2_{k}&=\begin{bmatrix} \overline{d}^{2\top}_{1,k} &   \dots & \overline{d}^{2\top}_{p,k}  \end{bmatrix}^\top, \quad \underline{d}^2_{k}=\begin{bmatrix} \underline{d}^{2\top}_{1,k} & \dots & \underline{d}^{2\top}_{p,k}  \end{bmatrix}^\top, \label{eq:dup2}\\
 \overline{d}^2_{i,k}&=\max \limits_{d_k \in \mathcal{D}_k} e_id_k , \ \ \underline{d}^2_{i,k}=\min \limits_{d_k \in \mathcal{D}_k} e_id_k, \forall i \in \{1,\dots p \}, \label{eq:dup2i}\\
 \overline{r}_k &= y_k-g_d(\underline{x}_k,\overline{x}_k)-Du_k-\underline{v},  \label{eq:rup} \\
  \underline{r}_k &= y_k-g_d(\overline{x}_k,\underline{x}_k)-Du_k-\overline{v},  \label{eq:rlow}
   \end{align}  
   with $J=H^\dagger$, $\tilde{H} \triangleq \begin{bmatrix} H^\top  -H^\top \end{bmatrix}^\top$,  $\tilde{r}_k \triangleq \begin{bmatrix} \overline{r}_k^\top & -\underline{r}_k^\top \end{bmatrix}^\top$, $\mathcal{D}_k \triangleq \{d_k | \tilde{H}d_k \leq \tilde{r}_k \}$, and 
$e_i \in \mathbb{R}^{1 \times p},  e_i(1,i)=1, \ e_i(1,j)=0, \forall j \neq i$. In the next sections, we will show that the choice of $J=H^\dagger$ and $f_d, g_d$ as decomposition functions of $f, g$ yields several desirable observer properties. The SISIO observer is summarized in Algorithm \ref{algorithm1}.

\begin{algorithm}[t] \small
\caption{Simultaneous Input and State Interval Observer}\label{algorithm1}
\begin{algorithmic}[1]
		\State Initialize: $J=H^\dagger$; $\text{maximal}(\mathcal{I}^x_0)=\overline{x}_0$; $\text{minimal}(\mathcal{I}^x_0)=\underline{x}_0$;
		\Statex  \hspace{0.1cm} Compute $J^{+}, J^{++}, G^{+}, G^{++}, L_{f_d}, L_{g_d}$ via Proposition \ref{prop:bounding} and \Statex  \hspace{0.1cm} Lemma \ref{lem:lip-dec}; Compute $g_d(\overline{x}_0,\underline{x}_0),g_d(\underline{x}_0,\overline{x}_0)$ via \eqref{eq:decompconstruct}; 
		\Statex \hspace{0.1cm} $K \triangleq (G^++G^{++})(J^++J^{++})$; $\Delta w = \overline{w}-\underline{w}$; $\Delta v = \overline{v}-\underline{v}$;
		\Statex \hspace{0.1cm} $\overline{r}_0 = y_0-g_d(\underline{x}_0,\overline{x}_0)-Du_0\hspace{-0.cm}-\hspace{-0.cm}\underline{v}$;
		\Statex \hspace{0.1cm} $\underline{r}_0 \hspace{-0.cm}= \hspace{-0.cm}y_0\hspace{-0.cm}-\hspace{-0.cm}g_d(\overline{x}_0,\underline{x}_0)\hspace{-0.cm}-\hspace{-0.cm}Du_0\hspace{-0.cm}-\hspace{-0.1cm}\overline{v}$; 
		\Statex \hspace{0.1cm} $\overline{d}^1_{0}\hspace{-0.cm} =\hspace{-0.cm}J^+ \overline{r}_0\hspace{-0.cm}-\hspace{-0.cm}J^{++}\underline{r}_0$; $\underline{d}^1_{0} \hspace{-0.cm}=\hspace{-0.cm}J^+ \underline{r}_0\hspace{-0.cm}-\hspace{-0.cm}J^{++}\overline{r}_0$; 
		\Statex \hspace{0.1cm}  $\Delta z \hspace{-0.cm}= \hspace{-0.cm} \Delta w\hspace{-0.cm}+\hspace{-0.cm}K \Delta v$; $\tilde{H} \triangleq \begin{bmatrix} H^\top  -H^\top \end{bmatrix}^\top$;
		\Statex \hspace{0.1cm} $\tilde{r}_0 \hspace{-0.cm}\triangleq \hspace{-0.cm}\begin{bmatrix} \overline{r}_0^\top & -\underline{r}_0^\top \end{bmatrix}^\top$; $\mathcal{D}_0 \hspace{-0.cm}\triangleq \hspace{-0.cm}\{d_0 | \tilde{H}d_0 \hspace{-0.cm}\leq\hspace{-0.cm} \tilde{r}_0 \}$;
		\Statex \hspace{0.1cm} $e_i \in \mathbb{R}^{1 \times p},  e_i(1,i) \hspace{-0.05cm}= \hspace{-0.05cm}1,  e_i(1,j)=0,\forall i,j \hspace{-0.05cm}\in \hspace{-0.05cm}\{1,\dots, p \}, j \hspace{-0.05cm}\neq \hspace{-0.05cm}i$;
		\Statex \hspace{0.1cm} $\forall i \in \{1,\dots p \}, \overline{d}^2_{i,0}=\max \limits_{d_0 \in \mathcal{D}_0} e_id_0$; $\underline{d}^2_{i,0}=\min \limits_{d_0 \in \mathcal{D}_0} e_id_0$;
		\Statex \hspace{0.1cm} $\delta^x_0=\| \overline{x}_0-\underline{x}_0 \|$; $\overline{d}_{0}=\min (\overline{d}^1_{0},\overline{d}^2_{0})$; $\underline{d}_{0}=\max (\underline{d}^1_{0},\underline{d}^2_{0})$; 
		\Statex \hspace{0.1cm} $\text{maximal}(\mathcal{I}^d_0)=\overline{d}_0$; $\text{minimal}(\mathcal{I}^d_0)=\underline{d}_0$; $\delta^d_0=\| \overline{d}_0-\underline{d}_0 \|$;
		\For {$k =1$ to $K$}

    	\LineComment{Estimation of ${x}_{k}$}
		\Statex \hspace{0.2cm} Compute $f_d(\overline{x}_{k-1},\underline{x}_{k-1}),f_d(\underline{x}_{k-1},\overline{x}_{k-1})$ via \eqref{eq:decompconstruct};
		\Statex \hspace{0.2cm} Compute $\overline{x}_{k},\underline{x}_{k}$ via \eqref{eq:xup} and \eqref{eq:xlow};
		\State \hspace{-0.2cm}$ 
		\delta^x_{k} =\mathcal{L}^k \delta_0^x + \|\Delta z\| \left(\frac{1-\mathcal{L}^k}{1-\mathcal{L}}\right)$;
		\State \hspace{-0.2cm}$\mathcal{I}^x_{k}=\{x \in \mathbb{R}^n : \underline{x}_k \leq x \leq \overline{x}_{k}\}$;
		\LineComment{Estimation of ${d}_{k}$}
		\Statex \hspace{0.2cm} Compute $g_d(\overline{x}_{k},\underline{x}_{k}),g_d(\underline{x}_{k},\overline{x}_{k})$ via \eqref{eq:decompconstruct};
		 \Statex \hspace{0.2cm} Compute $\overline{r}_k,\underline{r}_{k}$ via \eqref{eq:rup} and \eqref{eq:rlow}; $\tilde{r}_k \hspace{-0.1cm}\triangleq \hspace{-0.1cm}\begin{bmatrix} \overline{r}_k^\top & -\underline{r}_k^\top \end{bmatrix}^\top$; 
		 \Statex \hspace{0.2cm} Compute $\overline{d}^1_{k},\underline{d}^1_{k} $ via \eqref{eq:dup1}; $\mathcal{D}_k \hspace{-0.1cm}\triangleq \hspace{-0.1cm}\{d_k | \tilde{H}d_k \hspace{-0.1cm}\leq\hspace{-0.1cm} \tilde{r}_k \}$; 
		 \Statex \hspace{0.2cm} Compute $\overline{d}^2_{i,k},\underline{d}^2_{i,k} $ via \eqref{eq:dup2i} $\forall i \in \{1,\dots p \}$;
		 \Statex \hspace{0.2cm} Compute $\overline{d}_{k},\underline{d}_{k} $ via \eqref{eq:dup};
		\State \hspace{-0.2cm} $
		\delta^d_{k} = \|J^++J^{++}\| L_{g_d} \delta_k^x +\|(J^++J^{++})\Delta v \|$;
		\State \hspace{-0.2cm} $\mathcal{I}^d_{k}=\{d \in \mathbb{R}^p : \underline{d}_k \leq d \leq \overline{d}_{k}\}$;
		\EndFor

		\end{algorithmic}

\end{algorithm}      
      
\subsection{Correctness \moh{(Framer Property)} of Interval Estimates}
In the following, we show that 
the SISIO observer returns correct interval estimates 
in the sense that at each time step, the true states and unknown inputs are guaranteed to be within the estimated intervals 
given by \eqref{eq:xup}--\eqref{eq:dup}. \yong{
This is also known as the \emph{framer property}, e.g., in \cite{mazenc2013robust}.} \yongn{To increase readability, all proofs will be provided in the appendix.}
\begin{thm}[Correctness of the Interval Estimates]\label{thm:bounds}
Let $\underline{x}_0 \leq x_0 \leq \overline{x}_0$, where $\underline{x}_0$ and $\overline{x}_0$ are known. For the system \eqref{eq:system}, if Assumptions \ref{assumption:Hfull} and \ref{assumption:mix-monotone} hold, then the SISIO estimate intervals \eqref{eq:xup}--\eqref{eq:dup} with $J = H^\dagger$ and $f_d(\cdot,\cdot),  g_d(\cdot,\cdot)$ as decomposition functions of $f(\cdot), g(\cdot)$ at each step $k$ are \emph{correct}, i.e., the true states and unknown inputs are guaranteed to satisfy $\underline{d}_k \leq d_k \leq \overline{d}_k$ and $\underline{x}_k \leq x_k \leq \overline{x}_k$.
\end{thm}

\subsection{Boundedness of Interval Estimates and Observer Stability}
In this section we study the stability of SISIO, assuming that the decomposition functions can be obtained using \eqref{eq:decompconstruct}. We first derive a Lipschitz-like property for decomposition functions in Lemma \ref{lem:lip-dec}. Then, we derive several sufficient conditions for the stability of \yongn{SISIO} 
in Theorem \ref{thm:boundedness}.

\begin{lem}\label{lem:lip-dec}
Let $h(\cdot):\mathcal{D}_h \subseteq \mathbb{R}^n \rightarrow \mathbb{R}^m$ be a globally $L_h$-Lipschitz continuous and mixed monotone vector field and $h_d(\cdot,\cdot):\mathcal{D}_h \times \mathcal{D}_h \rightarrow \mathbb{R}^m $ be the decomposition function for $h$, constructed using \eqref{eq:decompconstruct}. Consider $\underline{x} \leq \overline{x}$, both in $\mathcal{D}_h$. Then $\|h_d(\overline{x},\underline{x}_k)-h_d(\underline{x}_k,\overline{x}_k)\| \leq L_{h_d}\|\overline{x}-\underline{x}\|$, where $L_{h_d} \triangleq L_h+2\|C_h\|$, with $C_h$ given in \eqref{eq:decompconstruct}.
\end{lem}
\begin{thm}[
Observer Stability] \label{thm:boundedness}
Consider the system \eqref{eq:system} and the SISIO observer \eqref{eq:xup}--\eqref{eq:dup}, and suppose that Assumption \ref{assumption:mix-lip} and all the assumptions and conditions in Theorem \ref{thm:bounds} hold and the decomposition functions $f_d, g_d$ are constructed using \eqref{eq:decompconstruct}. Then, the observer is stable, in the sense that at each time step $k$, interval width sequences $\{\|\Delta^d_k\| \triangleq \|\overline{d}_k-\underline{d}_k\|, \|\Delta^x_k\| \triangleq \|\overline{x}_{k}-\underline{x}_{k}\|\}_{k=0}^{\infty}$ are bounded, and consequently, interval input and state estimation errors $\{\|\tilde{d}_k\| \triangleq \max (\|d_k-\underline{d}_k\|,\|\overline{d}_k-{d}_k\|), \|\tilde{x}_{k}\| \triangleq \max (\|x_{k}-\underline{x}_{k}\|,\|\overline{x}_{k}-{x}_{k}\|) \}_{k=0}^{\infty}$ are also bounded, if \yong{\underline{either one}} of the following 
conditions hold:
\renewcommand{\theenumi}{\roman{enumi}}
\begin{enumerate}[(i)]
\item $\mathcal{L} \triangleq (L_{f_d}+L_{g_d}\|K\|) <1;$ \label{item:first}
\item \label{item:third}
$ \begin{bmatrix} F & 0 &0&0&0 \\ * & K^\top K & K^\top& K^\top&K^\top K \\ * & * & I & I & K \\ * & * & * &0 & K \\ * & * & * & * & 0 \end{bmatrix} \preceq 0;$
\item \label{item:second} There exist a positive definite matrix $P \succ 0$ and a positive semidefinite matrix $\Gamma \succeq 0$
in $\mathbb{R}^{n \times n}$ 
such that the following LMI condition \yongn{is 
satisfied:} 
\begin{align*}
\hspace{-2.2cm}\begin{bmatrix} P+\Gamma-I & 0 & P \\ 0 & \mathcal{L}I-P & 0 \\ P & 0 & P \end{bmatrix} \preceq 0, 
\end{align*}
\end{enumerate}
with $L_{f_d}$ and $L_{g_d}$ given in Lemma \ref{lem:lip-dec}, $K \triangleq (G^++G^{++})(J^++J^{++})$, $F\triangleq (L^2_{f_d}+L^2_{g_d}\lambda_{\max}(K^\top K)-1)I$ and $\lambda_{\max}(K^\top K)$ is the maximum eigenvalue of $K^\top K$.
\end{thm}

It is notable that examples of dynamic systems with only slight differences in their $G$ and $H$ matrices can be found, which 
only satisfy a subset of the three aforementioned 
conditions and do not satisfy the others. For instance, for the example system in Section \ref{sec:examples}, we found that it satisfies 
Conditions (i) and (ii) but does not satisfy 
Condition (iii). However, when we change the $G$ and $H$ matrices to $G=\begin{bmatrix}0 & 0.1 \\ 0.2 & -0.2 \end{bmatrix}$ and $H=\begin{bmatrix}0.1 & 0.3 \\ 0.5 &-0.7 \end{bmatrix}$, we observe that $\mathcal{L}=2.5212>1$, so 
Condition \ref{item:first} does not hold. In addition, 
Condition (ii) also does not hold, but 
Condition (iii) holds with $P=\begin{bmatrix}3.4352 & 0 \\ 0 &3.4352 \end{bmatrix}$ and $\Gamma=\begin{bmatrix}0.3402 & 0 \\ 0 &0.3402 \end{bmatrix}$. 
A search for a more general 
condition that 
\yongn{encompasses} all of these conditions is a subject of future work.


\moh{Finally, we will provide upper bounds for the interval widths and compute their \yong{steady-state values, if they exist}.}
\begin{lem}[Upper Bounds of the Interval Widths and their Convergence]\label{lem:convergence}
Consider the system \eqref{eq:system} and the SISIO observer \eqref{eq:xup}--\eqref{eq:dup}, and suppose that Assumptions \ref{assumption:Hfull}--\ref{assumption:mix-lip} hold. Then, at each time step $k$, there exist bounded and finite-valued upper bounds $\delta^x_{k}$ and $\delta^d_k$, for interval widths $\|\Delta^x_{k}\|$ and $\|\Delta^d_{k}\|$ respectively, which can be computed as follows:
\begin{align*}
\|\Delta^x_{k}\| &\leq \delta^x_{k} = \mathcal{L}^k \delta_0^x + \|\Delta z\| \left(\frac{1-\mathcal{L}^k}{1-\mathcal{L}}\right), \\
\|\Delta^d_{k}\| &\leq \delta^d_k = \|J^++J^{++}\| L_{g_d} \delta_k^x +\|(J^++J^{++})\Delta v \|, 
\end{align*}
with $L_{g_d}$ and $\mathcal{L}$ given in Lemma \ref{lem:lip-dec} and Theorem \ref{thm:boundedness}, respectively, $\Delta z \triangleq  \Delta w+K \Delta v$, $\Delta w \triangleq \overline{w}-\underline{w}$, $\Delta v \triangleq \overline{v}-\underline{v}$ and $K \triangleq (G^++G^{++})(J^++J^{++})$. Furthermore, if \moh{\yongn{Condition} \eqref{item:first} in Theorem \ref{thm:boundedness} holds}, then the upper bound sequences of the interval widths  converge to steady-state values as follows: 
\begin{align*}
\overline{\delta}^x &\triangleq \lim_{k \to \infty} \delta^x_{k} = \|\Delta z\|\frac{\mathcal{L}}{1-\mathcal{L}}, \quad \overline{\delta}^d \triangleq \lim_{k \to \infty} \delta^d_{k} =\mathcal{G}(\overline{\delta}^x),
\end{align*}
\yong{where $\mathcal{G}(x)\triangleq \|J^++J^{++}\| L_{g_d} x +\|(J^++J^{++})\Delta v \|$.}
\moh{\yong{On the other hand}, if \yongn{Conditions} \eqref{item:third} or \eqref{item:second} in Theorem \ref{thm:boundedness} hold, then the interval interval widths $\|\Delta^x_{k}\|$ and $\|\Delta^d_{k}\|$ are uniformly bounded by \yong{$\min\{\|\Delta^x_{0}\|,\Delta^P_0\}$ and $\min\{\mathcal{G}(\|\Delta^x_{0}\|),\mathcal{G}((\Delta^P_0)\}$,} 
respectively, \yong{with $\Delta^P_0 \triangleq \sqrt{\frac{(\Delta^x_{0})^\top P \Delta^x_{0}}{\lambda_{\min}(P)}}$ and $P$ being the}  solution for the LMI condition in \yongn{Condition} \eqref{item:second}.} 
\end{lem}
\vspace{-0.15cm}
\section{Illustrative Example} \label{sec:examples}
\begin{figure}[t]
\begin{center}
\includegraphics[scale=0.3,trim=25mm 0mm 5mm 5mm,clip]{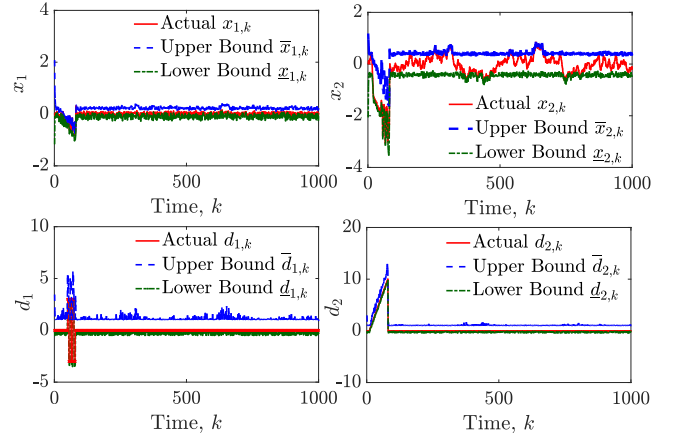}
\caption{Actual states and inputs, $x_{1,k}$, $x_{2,k}$, $d_{1,k}$, $d_{2,k}$, as well as their estimated maximal and minimal values, $\overline{x}_{1,k}$, $\underline{x}_{1,k}$, $\overline{x}_{2,k}$, $\underline{x}_{1,k}$, $\overline{d}_{1,k}$, $\underline{d}_{1,k}$, $\overline{d}_{2,k}$, $\underline{d}_{2,k}$. \label{fig:variances2}}
\end{center}
\vspace{-0.15cm}
\end{figure}

We consider the time-discretized dynamics of a nonlinear system in \cite{zheng2016design} with slight modifications to include measurement and process noise signals, and unknown inputs and with the following parameters (cf. \eqref{eq:system}):
$n=2$, $m=1$, $l=1$, $p=2$, $f(x_k)=\begin{bmatrix} f_1(x_k) & f_2(x_k)\end{bmatrix}^\top$, $f_1(x_k)=x_{2,k} + 0.25 \sin(0.1x_{1,k}x_{2,k})$, $f_2(x_k)=-0.2x_{2,k}-1.9\sin(0.01x_{1,k})$, $g(x_k)=x_{1,k}+0.526x_{2,k}-0.05x_{1,k}x_{2,k}$, $\mathcal{D}_f=\mathcal{D}_g=\begin{bmatrix} -5 & 5 \end{bmatrix}\times \begin{bmatrix} -15 & 15 \end{bmatrix}$, $B=\begin{bmatrix} 0 & 0.1 \end{bmatrix}^\top$, $D=0_{1 \times 2}$, $G=\begin{bmatrix} 0 & -0.1 \\ 0.2 & -0.2 \end{bmatrix}$, $H=\begin{bmatrix} -0.1 & 0.3 \\ 0.5 & -0.7 \end{bmatrix}$, $u_k=0.1\sin(k)+0.75\cos(0.25k)$, $\overline{v}=-\underline{v}=0.01$, $\overline{w}=-\underline{w}=\begin{bmatrix} 0.02 & 0.02 \end{bmatrix}^\top$, $\overline{x}_0=\begin{bmatrix} 2 & 1.1 \end{bmatrix}^\top$, $\underline{x}_0=\begin{bmatrix} -1.1 & -2 \end{bmatrix}^\top$, while the unknown input signals are depicted in Figure \ref{fig:variances2}. Note that rk$(H)=2$, thus Assumption \ref{assumption:Hfull} holds. Moreover, applying \cite[Theorem 1]{singh2018mesh}, we can compute finite-valued upper and lower bounds for partial derivatives of $f(\cdot)$ and $g(\cdot)$ as: $\begin{bmatrix}a^f_{11} & a^f_{12} \\ a^f_{21} & a^f_{22}  \end{bmatrix}=\begin{bmatrix} -0.25 & 0.99 \\ -0.0019 & -0.2\end{bmatrix}$, $\begin{bmatrix}b^f_{11} & b^f_{12} \\ b^f_{21} & b^f_{22}  \end{bmatrix}=\begin{bmatrix} 0.25 & 1.01 \\ 0.0019 & 0.2\end{bmatrix}$,  $\begin{bmatrix}a^g_{11} & a^g_{12} \end{bmatrix}=\begin{bmatrix} 0.75 & -0.224 \end{bmatrix}$, $\begin{bmatrix}b^g_{11} & b^g_{12} \end{bmatrix}=\begin{bmatrix} 1.25 & 0.1276\end{bmatrix}$. Hence, Assumption \ref{assumption:mix-monotone} is also satisfied by \cite[Theorem 1]{yang2019sufficient}). Therefore, we expect that the interval estimates are correct by Theorem \ref{thm:bounds} (i.e., the true states and unknown inputs are within the estimate intervals),  which can be verified from Figure \ref{fig:variances2} that depicts interval estimates as well as the true states and unknown inputs. 
\begin{figure}[t]
\begin{center}
\includegraphics[scale=0.325,trim=19mm 3mm 10mm 5mm,clip]{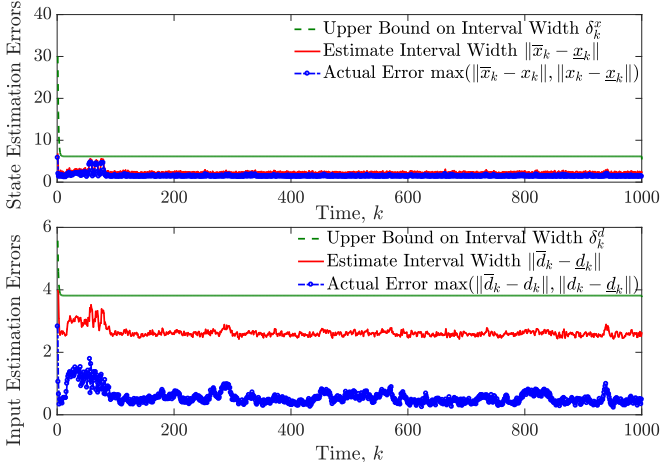}
\caption{Estimation errors, estimate interval widths and their upper bounds for the interval-valued estimates of states, $\|\tilde{x}_{k|k}\|$, \hspace{-0.05cm}$\|\Delta ^x_k\|$, \hspace{-0.05cm}$\delta^x_k$, and unknown inputs, $\|\tilde{d}_{k}\|$, $\|\Delta ^d_k\|$, $\delta^d_k$. \label{fig:variances3}}
\end{center}
\vspace{-0.2cm}
\end{figure}

Furthermore, $f(\cdot)$ and $g(\cdot)$ satisfy Assumption \ref{assumption:mix-lip} with $L_f=0.35$ and $L_g=0.74$. In addition, from \cite[(10)--(13)]{yang2019sufficient}), we obtain $C_f=\begin{bmatrix}  0.251 & 0 \\ 0.0029 & 0.201\end{bmatrix}$, $C_g=\begin{bmatrix} 0 & 0.225 \end{bmatrix}$ using \eqref{eq:decompconstruct}, which implies that $L_{f_d}=0.852$ and $L_{g_d}=1.19$ by Lemma \ref{lem:lip-dec}. Consequently, $\mathcal{L}=0.843$. Since all the required assumptions, including 
Condition \eqref{item:first} in Theorem \ref{thm:boundedness}, hold, we expect to obtain bounded and convergent interval estimate errors when applying our observer design procedure. This can be seen in Figure \ref{fig:variances3}, where at each time step, the actual error is less than or equal to the interval width, which in turn is less than or equal to the predicted upper bound for the interval width. Moreover, as expected, the upper bounds converge to some steady-state values. \yong{Note that, despite our best efforts, we were unable to find  interval-valued observers in the literature that simultaneously return both state and unknown input estimates for comparison with our results.} 
\vspace{-0.05cm}
\section{Conclusion} \label{sec:conclusion}
\vspace{-0.1cm}
In this paper, we proposed a simultaneous  input and state interval observer for mixed monotone Lipschitz nonlinear systems with unknown inputs and bounded noise. We proved that the proposed observer 
recursively outputs the state and unknown input interval-valued estimates that are guaranteed to include the true states and unknown inputs.
Moreover, several sufficient conditions for the stability of the observer and the boundedness of the interval widths were derived. 
Finally, we demonstrated the effectiveness of the proposed approach with an example. For future work, we seek to relax
the full-rank assumption for the direct feedthrough matrix \moh{and \yong{to find} necessary conditions for \yong{observer stability.}}
\vspace{-0.1cm}
\bibliographystyle{unsrturl}

\bibliography{biblio}

\begin{thebibliography}{10}

\bibitem{liu2011robust}
W.~Liu and I.~Hwang.
\newblock Robust estimation and fault detection and isolation algorithms for
  stochastic linear hybrid systems with unknown fault input.
\newblock {\em IET control theory \& applications}, 5(12):1353--1368, 2011.

\bibitem{yong2016tcps}
S.Z. Yong, M.~Zhu, and E.~Frazzoli.
\newblock Switching and data injection attacks on stochastic cyber-physical
  systems: {M}odeling, resilient estimation and attack mitigation.
\newblock {\em ACM Transactions on Cyber-Physical Systems}, 2(2):9, 2018.

\bibitem{yong2018simultaneous}
S.Z. Yong.
\newblock Simultaneous input and state set-valued observers with applications
  to attack-resilient estimation.
\newblock In {\em 2018 Annual American Control Conference (ACC)}, pages
  5167--5174. IEEE, 2018.

\bibitem{blanchini2012convex}
F.~Blanchini and M.~Sznaier.
\newblock A convex optimization approach to synthesizing bounded complexity
  $\ell^{\infty}$ filters.
\newblock {\em IEEE Transactions on Automatic Control}, 57(1):216--221, 2012.

\bibitem{jaulin2002nonlinear}
L.~Jaulin.
\newblock Nonlinear bounded-error state estimation of continuous-time systems.
\newblock {\em Automatica}, 38(6):1079--1082, 2002.

\bibitem{kieffer2004guaranteed}
M.~Kieffer and E.~Walter.
\newblock Guaranteed nonlinear state estimator for cooperative systems.
\newblock {\em Numerical algorithms}, 37(1-4):187--198, 2004.

\bibitem{moisan2007near}
M.~Moisan, O.~Bernard, and J-L. Gouz{\'e}.
\newblock Near optimal interval observers bundle for uncertain bioreactors.
\newblock In {\em European Control Conference (ECC)}, pages 5115--5122. IEEE,
  2007.

\bibitem{bernard2004closed}
O.~Bernard and J-L. Gouz{\'e}.
\newblock Closed loop observers bundle for uncertain biotechnological models.
\newblock {\em Journal of Process Control}, 14(7):765--774, 2004.

\bibitem{raissi2010interval}
T.~Ra{\"\i}ssi, G.~Videau, and A.~Zolghadri.
\newblock Interval observer design for consistency checks of nonlinear
  continuous-time systems.
\newblock {\em Automatica}, 46(3):518--527, 2010.

\bibitem{raissi2011interval}
T.~Ra{\"\i}ssi, D.~Efimov, and A.~Zolghadri.
\newblock Interval state estimation for a class of nonlinear systems.
\newblock {\em IEEE Transactions on Automatic Control}, 57(1):260--265, 2011.

\bibitem{mazenc2011interval}
F.~Mazenc and O.~Bernard.
\newblock Interval observers for linear time-invariant systems with
  disturbances.
\newblock {\em Automatica}, 47(1):140--147, 2011.

\bibitem{mazenc2013robust}
F.~Mazenc, T-N. Dinh, and S-I. Niculescu.
\newblock Robust interval observers and stabilization design for discrete-time
  systems with input and output.
\newblock {\em Automatica}, 49(11):3490--3497, 2013.

\bibitem{wang2015interval}
Y.~Wang, D-M. Bevly, and R.~Rajamani.
\newblock Interval observer design for {LPV} systems with parametric
  uncertainty.
\newblock {\em Automatica}, 60:79--85, 2015.

\bibitem{efimov2013interval}
D.~Efimov, T.~Ra{\"\i}ssi, S.~Chebotarev, and A.~Zolghadri.
\newblock Interval state observer for nonlinear time varying systems.
\newblock {\em Automatica}, 49(1):200--205, 2013.

\bibitem{zheng2016design}
G.~Zheng, D.~Efimov, and W.~Perruquetti.
\newblock Design of interval observer for a class of uncertain unobservable
  nonlinear systems.
\newblock {\em Automatica}, 63:167--174, 2016.

\bibitem{ellero2019unknown}
N.~Ellero, D.~Gucik-Derigny, and D.~Henry.
\newblock An unknown input interval observer for {LPV} systems under
  ${L}_2$-gain and ${L}_{\infty}$-gain criteria.
\newblock {\em Automatica}, 103:294--301, 2019.

\bibitem{khajenejad2019simultaneous}
M.~Khajenejad and S.Z. Yong.
\newblock Simultaneous input and state set-valued
  $\mathcal{{H}}_{\infty}$-observers for linear parameter-varying systems.
\newblock In {\em American Control Conference (ACC)}, pages 4521--4526. IEEE,
  2019.

\bibitem{khajenejadasimultaneous}
M.~Khajenejad and S.Z. Yong.
\newblock Simultaneous mode, input and state set-valued observers with
  applications to resilient estimation against sparse attacks.
\newblock In {\em Conference on Decision and Control (CDC)}, 2019.

\bibitem{yang2019sufficient}
L.~Yang, O.~Mickelin, and N.~Ozay.
\newblock On sufficient conditions for mixed monotonicity.
\newblock {\em IEEE Transactions on Automatic Control}, 64(12):5080--5085,
  2019.

\bibitem{coogan2015efficient}
S.~Coogan and M.~Arcak.
\newblock Efficient finite abstraction of mixed monotone systems.
\newblock In {\em Hybrid Systems: Computation and Control}, pages 58--67. ACM,
  2015.

\bibitem{yang2019tight}
L.~Yang and N.~Ozay.
\newblock Tight decomposition functions for mixed monotonicity.
\newblock In {\em Conference on Decision and Control (CDC)}, pages 5318--5322,
  2019.

\bibitem{gillijns2007unbiased}
S.~Gillijns and B.~De~Moor.
\newblock Unbiased minimum-variance input and state estimation for linear
  discrete-time systems with direct feedthrough.
\newblock {\em Automatica}, 43(5):934--937, 2007.

\bibitem{singh2018mesh}
K.R. Singh, Q.~Shen, and S.Z. Yong.
\newblock Mesh-based affine abstraction of nonlinear systems with tighter
  bounds.
\newblock In {\em Conference on Decision and Control (CDC)}, pages 3056--3061.
  IEEE, 2018.

\bibitem{khalil2002nonlinear}
H.K. Khalil.
\newblock Nonlinear systems.
\newblock {\em Upper Saddle River}, 2002.

\bibitem{delshad2016robust}
S.S Delshad, A.~Johansson, M.~Darouach, and T.~Gustafsson.
\newblock Robust state estimation and unknown inputs reconstruction for a class
  of nonlinear systems:\hspace{-0.05cm} {M}ultiobjective approach.
\newblock {\em \hspace{-0.02cm}Automatica}, \hspace{-0.02cm}64:1--7, 2016.

\end{thebibliography}

\section*{Appendix: Proofs} 
\label{subsec:thmproof}
\vspace{-0.1cm}
\subsection{Proof of Theorem \ref{thm:bounds}}
First, note that for $r_k\triangleq H d_k = y_k-g(x_k)-Du_k-v_k$, we can obtain $\underline{r}_k \leq r_k=H d_k \leq \overline{r}_k$ by Assumption \ref{assumption:mix-monotone} and the fact that decomposition functions are monotone increasing in their first argument and decreasing in their second (cf. Definition \ref{defn:mixed-monotone}). By left multiplying the above inequalities by $J=H^\dagger$ and from Assumption \ref{assumption:Hfull} and Proposition \ref{prop:bounding}, 
we can conclude that $\underline{d}^1_k \leq d_k \leq \overline{d}^1_k$. Moreover,
since $\underline{r}_k \leq r_k=H d_k \leq \overline{r}_k$ can be rearranged as ${d_k \in \mathcal{D}_k \triangleq \{d \in \mathbb{R}^p\, |}\,\tilde{H} d \leq \tilde{r}_k{\}}$, the linear programs \eqref{eq:dup2i} 
yield the tightest maximal and minimal elements of $\mathcal{I}^d_k$ that enclose {$\mathcal{D}_k$}, i.e.,  $\underline{d}^2_k \leq d_k \leq \overline{d}^2_k$. Combining this and 
 $\underline{d}^1_k \leq d_k \leq \overline{d}^1_k$, we obtain 
 $\underline{d}_k \leq d_k \leq \overline{d}_k$. 
 
 Similarly, since Assumption \ref{assumption:mix-monotone} holds, by applying the fact that decomposition functions are monotone increasing in their first argument and decreasing in their second, as well as Proposition \ref{prop:bounding} to \eqref{eq:system}, we obtain $\underline{x}_{k} \leq x_{k} \leq \overline{x}_{k}$ with $\underline{x}_{k}$ and $\overline{x}_{k}$ given in \eqref{eq:xup} and \eqref{eq:xlow}.
\qed

\vspace{-0.1cm}
\subsection{Proof of Lemma \ref{lem:lip-dec}}
Starting from \eqref{eq:decompconstruct},
\begin{align}
 f_d(\overline{x},\underline{x})=f(x_1)+C_f(\overline{x}-\underline{x})\label{eq:fd}, \\
  f_d(\underline{x},\overline{x})=f(x_2)+C_f(\underline{x}-\overline{x}),\label{eq:fd2}
  \end{align}
   where {$\forall i \in \{1\dots n\}$,} $x_{1,{i}}$ and $x_{2,\moh{i}}$ are either $\overline{x}{_i}$, or $\underline{x{_i}}$, depending on the case (cf. \cite[Theorem 1; (10)--(13)]{yang2019sufficient}). Moreover, $\underline{x}\leq \overline{x}$ and $ \underline{x}\leq x_1,x_2 \leq \overline{x}.$ This implies that
\begin{align}\label{eq:order}
 -(\overline{x}-\underline{x})\hspace{-0.05cm} \leq \hspace{-0.05cm} x_1\hspace{-0.1cm} -\hspace{-0.1cm} x_2 \leq \overline{x}-\underline{x}\hspace{-0.1cm} \implies \hspace{-0.2cm} \|x_1-x_2\|\hspace{-0.1cm}  \leq \hspace{-0.1cm} \| \overline{x}-\underline{x}\|. 
 \end{align} 
On the other hand, from \eqref{eq:fd} and \eqref{eq:fd2}, $$f_d(\overline{x},\underline{x})- f_d(\underline{x},\overline{x})=f(x_1)-f(x_2)+2C_f(\overline{x}-\underline{x}).$$ Then, applying 
triangle inequality and by the Lipschitz continuity of $f$, we obtain 
 \begin{align} \label{eq:fd3}
  \hspace{-0.15cm}\|f_d(\overline{x},\underline{x})\hspace{-0.1cm}-\hspace{-0.1cm}f_d(\underline{x},\overline{x})\|\hspace{-0.05cm} \leq \hspace{-0.05cm}L_f\|x_1\hspace{-0.05cm}-\hspace{-0.05cm} x_2\|\hspace{-0.05cm}+\hspace{-0.05cm}2\|C_f\|\|(\overline{x}\hspace{-0.05cm}-\hspace{-0.05cm}\underline{x})\|.
  \end{align}
  Combining \eqref{eq:order} and \eqref{eq:fd3} 
  yields the result. 
\qed
\subsection{Proof of Theorem \ref{thm:boundedness}}
From \eqref{eq:xup} and \eqref{eq:xlow}, we obtain 
\begin{align}\label{eq:deltax}
\Delta^x_{k+1}=\Delta f^x_k+(G^++G^{++})\Delta^d_k+\Delta w,
\end{align}
where $\Delta^x_{k} \triangleq \overline{x}_k-\underline{x}_k$, $\Delta^d_k \triangleq \overline{d}_k-\underline{d}_k$, $\Delta f^x_k \triangleq f_d(\overline{x}_k,\underline{x}_k)-f_d(\underline{x}_k,\overline{x}_k)$ and $\Delta w \triangleq \overline{w}-\underline{w}$. Moreover, from \eqref{eq:dup} and \eqref{eq:dup1}, 
\begin{align} \label{eq:deltad}
\Delta^d_k \leq \Delta^{d,1}_k =  (J^++J^{++}) \Delta^r_k,
\end{align} 
where $\Delta^{d,1}_k \triangleq \overline{d}^1_k-\underline{d}^1_k$ and $\Delta^r_k \triangleq \overline{r}_k-\underline{r}_k$, while 
from the definitions of $\overline{r}_k$ and $\underline{r}_k$ in \eqref{eq:rup}--\eqref{eq:rlow}, we have 
\begin{align} \label{eq:deltar}
\Delta^r_k = \Delta g^x_k+\Delta v,
\end{align}
where $\Delta g^x_k \triangleq g_d(\overline{x}_k,\underline{x}_k)-g_d(\underline{x}_k,\overline{x}_k)$ and $\Delta v \triangleq \overline{v}-\underline{v}$. 
Combining \eqref{eq:deltax}--\eqref{eq:deltar} and using the fact that $G^+,G^{++},J^+,J^{++}\geq 0$ and Proposition \ref{prop:bounding}, we obtain
\begin{align}\label{eq:deltax2}
\Delta^x_{k+1} \le \Delta h^x_k\hspace{-0.1cm}+\hspace{-0.1cm}\Delta z,
\end{align}
where $K \triangleq (G^++G^{++})(J^++J^{++})$, $\Delta h^x_k \triangleq \Delta f^x_k\hspace{-0.1cm}+\hspace{-0.1cm}K\Delta g^x_k$ and $\Delta z \triangleq \Delta w+K \Delta v$. Now, consider the following dynamical system
\begin{align} \label{tdynm}
\Delta^s_{k+1} = \Delta h^s_k\hspace{-0.1cm}+\hspace{-0.1cm}\Delta z,
\end{align}
where $\Delta^s_{k} \in \mathbb{R}^n$ and $\Delta^s_{0}=\Delta^x_{0}$. 
By {non-negativity of $\Delta^x_k$ and} \emph{Comparison Lemma} \cite[Lemma 3.4]{khalil2002nonlinear}, $0 \leq \Delta^x_k \leq \Delta^s_{k}, \forall k\geq 0$. {So, boundedness of $\{\Delta^s_k\}_{k=0}^{\infty}$ (shown below) implies boundedness of $\{\Delta^x_k\}_{k=0}^{\infty}$}. 

\noindent\textbf{
Condition (i):}
 Since Assumption \ref{assumption:mix-lip} holds, the application of triangle inequality to \eqref{eq:deltax2} yields 
\begin{align} \label{deltaxdynm}
\|\Delta^x_{k+1}\| \leq \mathcal{L}\|\Delta^x_k\|+\|\Delta z\|,
\end{align}
where $\mathcal{L} = L_{f_d}+L_{g_d}\|K\|$, with $L_{f_d}$ and $L_{g_d}$ given in Lemma \ref{lem:lip-dec}. Since $\mathcal{L}<1$ (by 
Condition (i)), 
by \emph{Comparison Lemma} \cite[Lemma 3.4]{khalil2002nonlinear}, 
the $\{\|\Delta^x_k\|\}_{k=0}^{\infty}$ is bounded. Therefore, the interval width dynamics is stable.\\[-0.25cm] 

\noindent\textbf{
Condition (ii):}
To show that 
Condition (\ref{item:third}) implies stability, 
consider a candidate Lyapunov function $V_k=\Delta^{s\top}_k\Delta^s_k$ for \eqref{tdynm} that can be shown to satisfy $\Delta V_k \triangleq V_{k+1}-V_k \leq 0$ under 
Condition (\ref{item:third}) as follows: 

\hspace{0.1cm} \vspace{-0.55cm}
\begin{align*}
&\Delta V_k \triangleq V_{k+1}-V_k\\
&=\Delta f^{s\top}_k \hspace{-0.05cm}\Delta f^s_k \hspace{-0.05cm}+\hspace{-0.05cm} \Delta g^{s\top}_k K^\top \hspace{-0.05cm} K \Delta g^s_k \hspace{-0.05cm}+\hspace{-0.05cm} \Delta v^\top K^\top \hspace{-0.05cm} K \Delta v \hspace{-0.05cm}+\hspace{-0.05cm} \Delta w^\top \hspace{-0.05cm} \Delta w\\
& \ \ -\Delta^{s\top}_k\Delta^s_k+2(\Delta f^{s\top}_k K \Delta g^s_k+\Delta f^{s\top}K\Delta v+\Delta f^{s\top}\Delta w\\
& \ \ +\Delta g^{s\top}K^\top K \Delta v+\Delta g^{s\top}K^\top \Delta w+\Delta v^\top K^\top \Delta w)\\
& \leq (L^2_{f_d}+\lambda_{\max}(K^\top K)L^2_{g_d}-1)\Delta^{s\top}_k\Delta^s_k+\Delta v^\top K^\top K \Delta v\\
& \ \ +\Delta w^\top \Delta w+2(\Delta f^{s\top}_k K \Delta g^s_k+\Delta f^{s\top}K\Delta v+\Delta f^{s\top}\Delta w\\
& \ \ +\Delta g^{s\top}K^\top K \Delta v+\Delta g^{s\top}K^\top \Delta w+\Delta v^\top K^\top \Delta w)\\
&= 
\begin {bmatrix} \Delta^s_k \\ \Delta v \\ \Delta w \\ \Delta f^s_k \\ \Delta g^s_k \end{bmatrix}^\top \begin{bmatrix} F & 0 &0&0&0 \\ * & K^\top K & K^\top& K^\top&K^\top K \\ * & * & I & I & K \\ * & * & * &0 & K \\ * & * & * & * & 0 \end{bmatrix}\begin {bmatrix} \Delta^s_k \\ \Delta v \\ \Delta w \\ \Delta f^s_k \\ \Delta g^s_k \end{bmatrix}\le 0,
\end{align*}
where the first inequality holds since $\Delta f^{s\top}_k\Delta f^s_k=\|\Delta f^s_k\|^2 \leq L^2_{f_d} \|\Delta^s_k\|^2$ by Lemma \ref{lem:lip-dec} and $\Delta g^{s\top}_kK^\top K\Delta g^s_k\leq \lambda_{\max}(K^\top K)\Delta g^{s\top}_k\Delta g^s_k=\lambda_{\max}(K^\top K)\|\Delta g^s_k\|^2 \leq L^2_{g_d}\lambda_{\max}(K^\top K) \|\Delta^s_k\|^2$ by using the \emph{Rayleigh Quotient} and Lemma \ref{lem:lip-dec}, and  the last inequality holds by 
Condition (\ref{item:third}). Thus, $\Delta_k^s$ is bounded and so is $\Delta_k^x$ by the Comparison Lemma  (i.e., the dynamics of $\Delta^x_k$ is stable).\\[-0.25cm] 

\noindent\textbf{
Condition (iii):}
Similarly, we consider a {candidate} 
Lyapunov function $V_k=\Delta^{s\top}_k P \Delta^{s}_k$, where $P \succ 0$, which can be shown to satisfy $\Delta V_k \triangleq V_{k+1}-V_k \leq 0$ under 
Condition (\ref{item:second}). To show this, 
note that  
$\Delta h^{s\top}_k\Lambda \Delta h^s_k \le \Delta h^{s\top}_k \Delta h^s_k \leq \mathcal{L}^2\Delta^{s\top}_k \Delta^s_k$, where the inequalities hold by {choosing $\Gamma$ such that} $\Gamma \triangleq I -\Lambda \succeq 0$ and Lemma \ref{lem:lip-dec}, respectively. Consequently, $\mathcal{L}^2\Delta^{s\top}_k \Delta^s_k-\Delta h^{s\top}_k\Lambda \Delta h^s_k \ge 0$. Then, inspired by a simplifying trick used in \cite[Proof of Theorem 1]{delshad2016robust} to satisfy $\Delta V_k \le 0$, it suffices to guarantee that $\tilde{V}_k \triangleq \Delta V_k +\mathcal{L}^2\Delta^{s\top}_k \Delta^s_k-\Delta h^{s\top}_k\Lambda \Delta h^s_k = \Delta V_k +\mathcal{L}^2\Delta^{s\top}_k \Delta^s_k-\Delta h^{s\top}_k(I-{\Gamma}) \Delta h^s_k \le 0$, where $\tilde{V}_k$ is also given by
\begin{align*}
\tilde{V}_k &=\Delta h^{s\top}_k P \Delta h^{s}_k \hspace{-0.1cm}+\hspace{-0.1cm}\Delta z^\top P \Delta z\hspace{-0.1cm}+\hspace{-0.1cm}2\Delta z^\top P\Delta h^s_k\hspace{-0.1cm}-\hspace{-0.1cm}\Delta^{s\top}_k P \Delta^s_k\\
&\ \ +\mathcal{L}^2\Delta^{s\top}_k \Delta^s_k-\Delta h^{s\top}_k (I-{\Gamma}) \Delta h^s_k\\
&= \Delta h^{s\top}_k(P+\Gamma-I)\Delta h^s_k+\Delta^{s\top}_k(\mathcal{L}^2I-P) \Delta^s_k\\
& \ \ +\Delta z^\top P \Delta z+2\Delta z^\top P \Delta h^s_k \\
&=\begin{bmatrix} \Delta h^s_k \\ \Delta^s_k \\ \Delta z \end{bmatrix}^\top \begin{bmatrix} P+\Gamma-I & 0 & P \\ 0 & \mathcal{L}^2I-P & 0 \\ P & 0 & P \end{bmatrix} \begin{bmatrix} \Delta h^s_k \\ \Delta^s_k \\ \Delta z \end{bmatrix} \le 0, 
\end{align*}
which along with $\Gamma \succeq 0 $ is equivalent to 
Condition  \eqref{item:second}. Finally, since $\Delta V_k \le 0$, $\Delta_k^s$ is bounded and so are the interval width sequences $\{\Delta_k^x\}_{k=0}^\infty$ by the Comparison Lemma (i.e., the dynamics of $\Delta^x_k$ is stable).
\qed

\subsection{Proof of \moh{Lemma} \ref{lem:convergence}}
Applying \eqref{deltaxdynm} repeatedly, we have
\begin{align*}
\|\Delta^x_{k}\|\leq \mathcal{L}^k\|\Delta^x_0\|+\textstyle{\sum}_{i=0}^{k-1}\mathcal{L}^{k-i}\|\Delta z\|= \mathcal{L}^k \delta_0^x + \|\Delta z\| \frac{1-\mathcal{L}^k}{1-\mathcal{L}}. 
\end{align*}
Similarly, by applying Lemma \ref{lem:lip-dec} and triangle inequality to \eqref{eq:deltad} and \eqref{eq:deltar}, we obtain the upper bound $\|\Delta^d_k\|$. \moh{If} $\mathcal{L}<1$, \moh{then} taking the limit of $k$ to $\infty$, \moh{returns} $\overline{\delta}^x$ and $\overline{\delta}^d$. \moh{The \yong{rest of the results follow from the}
non-increasing Lyapunov functions defined in the proof of Theorem \ref{thm:boundedness}, as well as the fact that $\lambda_{\min}(A)\|x\|^2 \leq x^\top A x, \forall x \in \mathbb{R}^n,A \in \mathbb{R}^{n \times n}$.}
\qed
\end{document}